# Electrical double layer: revisit based on boundary conditions

Jong U. Kim[*]

Department of Electrical and Computer Engineering, Texas A&M University

College Station, TX 77843-3128, USA

Abstract

The electrical double layer at infinite flat solid surface was discussed with respect to the triple layer model. Charge density at the outer Helmholtz plane is assumed to be zero usually. However, the assumption causes the position of the outer Helmholtz plane to be obscure. The charge density was calculated and the condition that the charge density is negligible was discussed.

*Keywords*: Electrical double layer; Stern layer; outer Helmholtz plane; triple layer model

---

[*] Corresponding author. *E-mail address*: jongkim@ee.tamu.edu

## 1. Introduction

Understanding phenomena near the interface of solid and an aqueous solution is of importance in electrokinetics, microfluidics, colloidal dispersion, and electrochemistry. When solid surface contacts an aqueous solution including electrolyte, the solid surface becomes charged due to the difference of electron (or ion) affinities between the solid surface and the solution or the ionization of surface groups. In addition, the surface charges cause a special structure at the interface, so called the electrical double layer (EDL) [1,2].

Usually Gouy-Chapman-Stern model (GCS) is widely used to describe the EDL. The GCS model consists of two layers; Stern layer (SL) and diffuse layer (DL). The SL is the region next to the surface, and ions in the SL are bound near the surface due to specially-adsorbing and Coulomb interactions. The DL is the region next to the SL and ions in the DL can move freely in any direction. The SL has two planes; the inner Helmholtz plane (IHP) and the outer Helmholtz plane (OHP) as shown in Figure 1. In general, the surface charge density and the charge density at the IHP are predicted in thermodynamic approach or site-binding models [1-4]. However, the charge density at the OHP is assumed to be zero without any consideration.

In this letter, we will evaluate the charge density at the OHP as a function of the sum of the surface charge density and the charge density at the IHP and discuss the condition that the charge density at the OHP is negligible.

## 2. Electrical double layer

Let's consider infinite flat solid surface in contact with an aqueous solution including electrolyte. The surface is uniform. Figure 1 shows a schematic diagram of the EDL. To easily understand the structure of the EDL, we introduce three types of ions in the solution; potential-determining, specifically-adsorbed and indifferent ions [5]. Potential-determining ions are adsorbed at the surface directly. Their equilibrium distribution between the surface and the solution determines the surface potential relative to potential in bulk solution. The adsorbed potential-determining ions form the surface charge density. Indifferent ions are affected by Coulomb force of the surface charge. Thus, they are repelled by the same sign surface charges while they are attracted by the opposite sign. Specifically-adsorbed ions are strongly interacted with the surface through all interactions other than purely Coulomb force. In the triple layer model, the IHP is located at the center of specifically-adsorbed ions and the OHP is located at the center of indifferent ions [1]. Here we consider only triple layer model which has the specific adsorbed ions. However, our conclusion may be applicable to other EDL models.

In the triple layer model, potentials in the SL consisting of the region 1 and the region 2 shown in Fig. 1 satisfy

$$\frac{\partial^2 \psi}{\partial x^2} = 0 \quad . \tag{1}$$

If we apply the potential-related boundary conditions at the surface, the IHP and the OHP, the potential profiles in the SL are

$$\psi_1(x) = \psi_0 + (\psi_i - \psi_0) x/\beta \quad \text{at} \quad 0 \leq x \leq \beta \quad \text{(region 1)} \tag{2}$$

and

$$\psi_2(x) = \psi_i + (\zeta - \psi_i)\frac{x - \beta}{\delta - \beta} \quad \text{at} \quad \beta \leq x \leq \delta \quad \text{(region 2)} \quad , \tag{3}$$

where $\psi_0$ and $\psi_i$ are, respectively, the surface potential and the potential at the IHP, $\beta$ and $\delta$ are, respectively, the positions of the IHP and of the OHP from the surface, and $\zeta$ is the potential at the OHP.

The above potential profiles in the SL are obtained through boundary condition based on potential but they also satisfy boundary conditions based on electric displacement. Usually, it is assumed in the triple layer model that the charge density at the OHP is zero. However, the position of the OHP is obscure if there is no charge at the OHP. Therefore, the boundary condition at the OHP need to include the charge density at the OHP. The boundary conditions based on electric displacement at the surface, the IHP and the OHP are, respectively, given by

$$\sigma_0 = -\varepsilon_0 \varepsilon_1 \frac{\partial \psi_1}{\partial x}\bigg|_{x=0} \quad \text{at } x=0 \text{ (the surface)},\tag{4}$$

$$-\varepsilon_0 \varepsilon_1 \frac{\partial \psi_1}{\partial x}\bigg|_{x=\beta} + \sigma_i = -\varepsilon_0 \varepsilon_2 \frac{\partial \psi_2}{\partial x}\bigg|_{x=\beta} \quad \text{at } x=\beta \text{ (the IHP)},\tag{5}$$

and

$$-\varepsilon_0 \varepsilon_2 \frac{\partial \psi_2}{\partial x}\bigg|_{x=\delta} + \sigma_{OHP} = -\varepsilon_0 \varepsilon_b \frac{\partial \psi_{\text{diffuse}}}{\partial x}\bigg|_{x=\delta} \quad \text{at } x=\delta \text{ (the OHP)},\tag{6}$$

where $\sigma_0$, $\sigma_i$ and $\sigma_{OHP}$ are the charge densities at the surface, the IHP and the OHP, respectively, and $\varepsilon_1$ and $\varepsilon_2$ are the dielectric constants in the region 1 and the region 2, respectively. These dielectric constants are generally different. The dielectric constant in the region 1 is discussed by Sverjensky [4]. The dielectric constant in the DL is thought of as the dielectric constant of bulk water (77.78 at 300K). Summing Eqs. (4) through (6) gives

$$\sigma_0 + \sigma_i + \sigma_{OHP} = -\varepsilon_0 \varepsilon_b \frac{\partial \psi_{\text{diffuse}}}{\partial x}\bigg|_{x=\delta}.\tag{7}$$

Since the virtual charge density in the DL is defined as

$$\sigma_d = \varepsilon_0 \varepsilon_b \frac{\partial \psi_{\text{diffuse}}}{\partial x}\bigg|_{x=\delta},\tag{8}$$

Eq. (7) is rewritten as

$$\sigma_0 + \sigma_i + \sigma_{OHP} + \sigma_d = 0.\tag{9}$$

Equation (9) is electro-neutrality condition and is the same as that in Ref. [1] except including $\sigma_{OHP}$.

For simplicity of notation, let's introduce effective surface charge density (ESCD), saying $\sigma_s = \sigma_0 + \sigma_i$, and charge density ratio defined by $\sigma_{OHP} = -\gamma \sigma_s$ where $0 < \gamma \leq 1$ [6]. Since the OHP is not movable in the normal direction of the surface, the sum of Coulomb force per area acting on the charges at the OHP due to the effective surface charge, electrostriction pressure of the water in the region 2, and electrostriction pressure of the solution in the DL has to be zero. The Coulomb force acting on the charges at the OHP is

$$F_C = -\frac{\gamma \sigma_s^2}{4\pi \varepsilon_0 \varepsilon_2} \iint dy' dz' \iint dy dz \frac{\delta \mathbf{i} + (y-y')\mathbf{j} + (z-z')\mathbf{k}}{\left[\delta^2 + (y-y')^2 + (z-z')^2\right]^{3/2}} \quad , \tag{10}$$

where $\mathbf{i}$, $\mathbf{j}$ and $\mathbf{k}$ are the unit vectors in x, y and z directions, respectively. Integrating the right hand side of Eq. (10) yields

$$\frac{F_C}{A} = -\frac{\gamma \sigma_s^2}{2\varepsilon_0 \varepsilon_2} \quad . \tag{11}$$

This force acts on the charges at the OHP in the negative x direction. That is, the water in the region 2 is compressed by the charges at the OHP.

There are two kinds of electrostriction pressure [7]. One is suitable for constant volume system and the other is suitable for constant chemical potential system. If water molecules freely enter or leave the region 2, the latter should be used; otherwise, the former should. Since the position of the OHP is fixed, the volume of the water in the region 2 is constant. Thus, electrostriction pressure for constant volume system is appropriate for the EDL. Electrostriction pressures of fluid at constant volume system, $P$, satisfies [7]

$$dP = -\varepsilon_0 \left[ \rho \left( \frac{\partial \varepsilon}{\partial \rho} \right)_{E,T} - (\varepsilon - 1) \right] E \, dE \quad , \tag{12}$$

where $\varepsilon$ is the dielectric constant of fluid, $\rho$ is density, and $E$ is electric field. The dielectric constant of water as a function of electric field strength $E$ is given by [8]

$$\varepsilon = n^2 + \frac{7\rho \mu^2 (n^2+2)}{3\varepsilon_0 \sqrt{73} E} L\left( \frac{\sqrt{73} E \mu (n^2+2)}{6 k_B T} \right) \quad , \tag{13}$$

where $\mu$ is electric dipole of a single water molecule (2.02 Debye units), $n$ is the optical refractive index of water (1.33 at 300K), and $L(x)$ is the Langevin function give by $L(x) = \coth(x) - 1/x$. Combining Eq. (13) with Eq. (12) and integrating both sides from zero electric field to electric field in the region 2 yield

$$P = P_0 + \frac{(n^2-1)\sigma_s^2}{2\varepsilon_0 \varepsilon_2^2} \qquad (14)$$

where $P_0$ is the electric field-free pressure. Here the electric field in the region 2 is obtained as $E_2 = \sigma_s / \varepsilon_0 \varepsilon_2$ by use of Eqs. (4) and (5).

Since there is no pressure-driven flow in the DL, the pressure in the DL is constant. In addition, it is field-free pressure in bulk solution. Therefore, the electrostriction pressure in the DL is equal to the field-free pressure $P_0$. In order that the OHP does not move in the normal direction of the surface, the pressure of the water in region 2 is equal to the sum of the Coulomb force per area and the pressure of bulk solution;

$$P_0 + \frac{(n^2-1)\sigma_s^2}{2\varepsilon_0 \varepsilon_2^2} = \frac{\gamma \sigma_s^2}{2\varepsilon_0 \varepsilon_2} + P_0 \quad, \qquad (15)$$

or

$$\gamma = \frac{-\sigma_{OHP}}{\sigma_0 + \sigma_i} = \frac{n^2-1}{\varepsilon_2} \quad . \qquad (16)$$

Figure 2 shows the dielectric constant of the water in the region 2 as a function of the ESCD. The dielectric constant is numerically calculated by using Eq. (13) and $E = |\sigma_s|/\varepsilon_0 \varepsilon$. It is shown in Fig. 2 that the dielectric constant in the region 2 decreases with increasing the ESCD and that it is approximately the dielectric constant of bulk water at the ESCD of less than 10 µm/cm². Figure 3 shows the dependence of the charge density ratio $\gamma$ on the ESCD. It is shown in Fig. 3 that the charge density ratio is approximately 0.01 at the ESCD of smaller than 10 µC/cm². However, it is also shown that the charge density at the OHP cannot be ignored at the ESCD of bigger than 30 $\mu C/cm^2$. Therefore, only when the ESCD is smaller than 10 $\mu C/cm^2$, the charge density at the OHP is negligible. That is, the existent triple layer model is available when the ESCD is smaller than 10 $\mu C/cm^2$.

## 3. Summary

In summary, the charge density at the OHP was calculated. It was shown that the charge density at the OHP is negligible at the effective surface charge density of smaller than 10 $\mu C/cm^2$.

## Acknowledgement

The author would like to acknowledge the support of Ebensbeger/Fouraker Graduate Fellowship.

**Figure Captions**

Fig. 1. Schematic diagrams of the electrical double layers with respect to the triple layer model. The numbers 1 and 2 represent, respectively, region 1 and region 2, and the long-dashed line represents potential profile when the surface charge is negative.

Fig. 2. Dielectric constant of the water in the region 2 as a function of the absolute value of the effective surface charge density.

Fig. 3. Charge density ratio $\gamma$ as a function of the absolute value of the effective surface charge density.

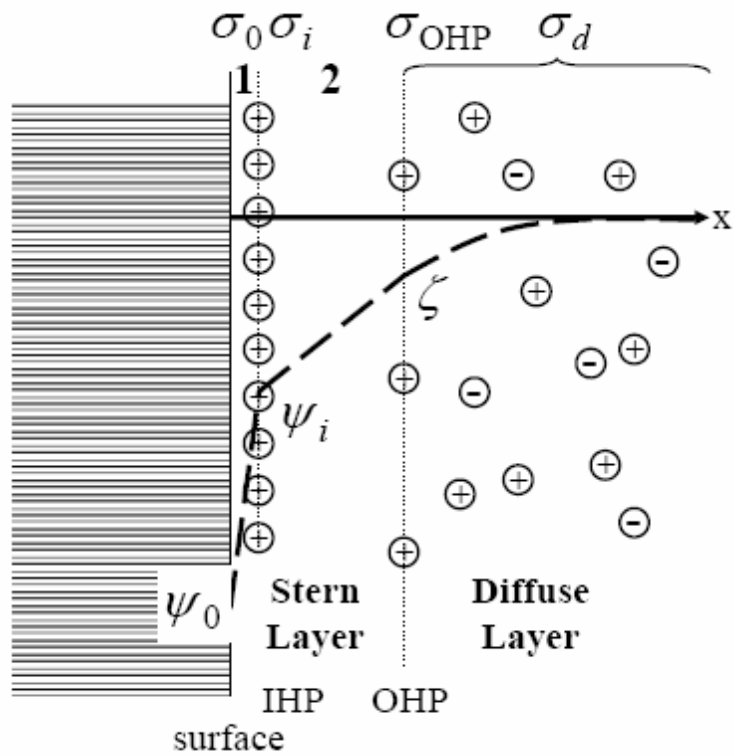

Fig. 1

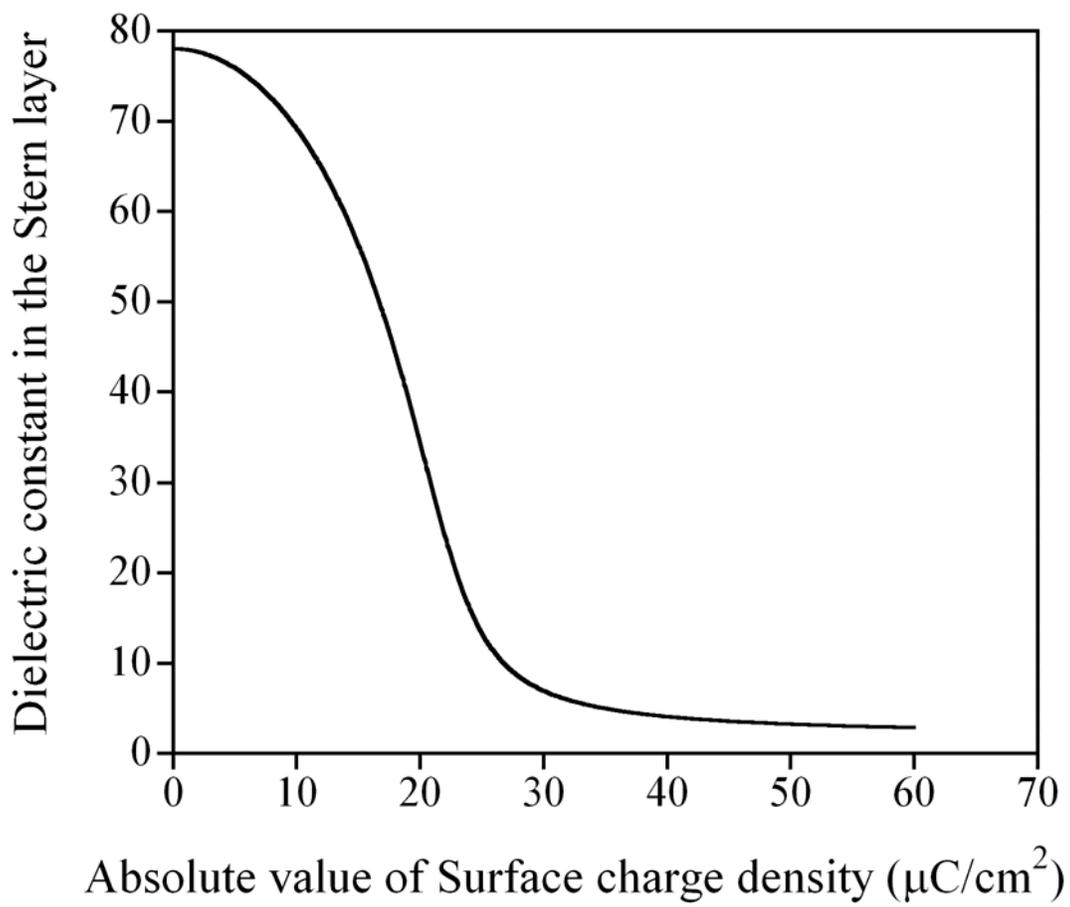

Fig. 2

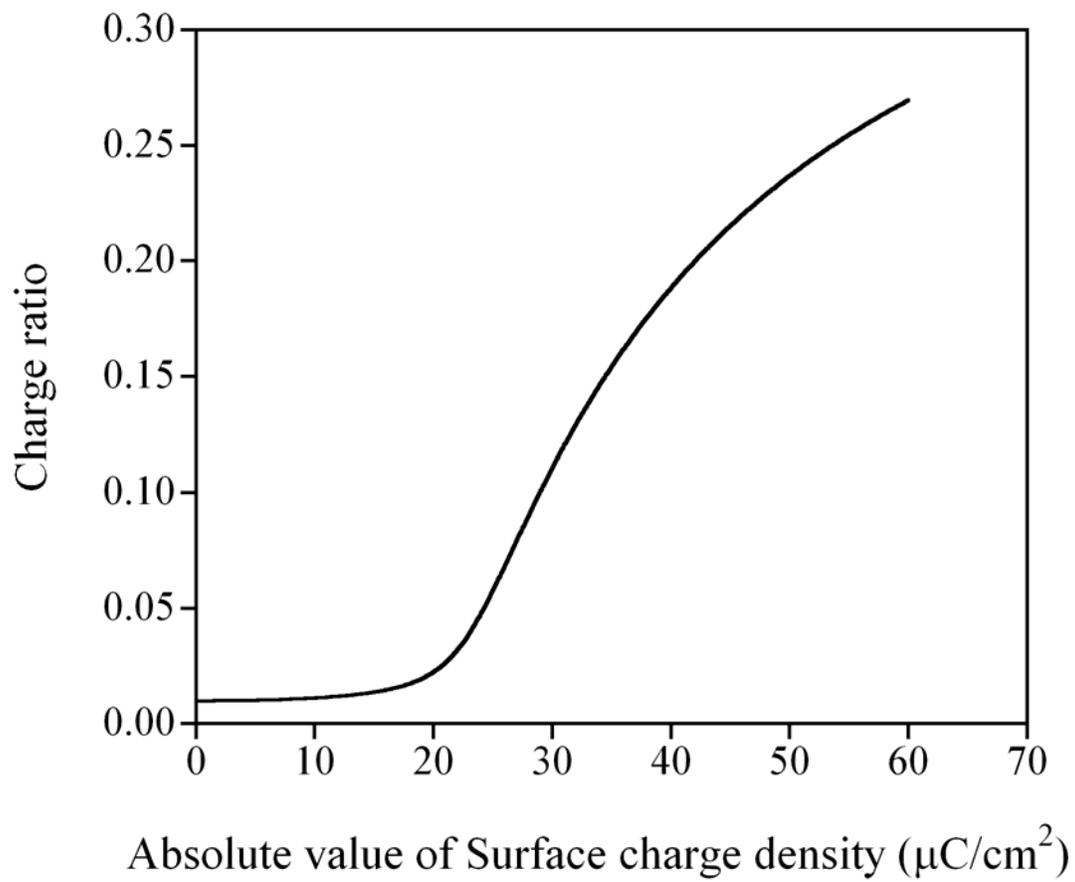

Fig. 3